\documentclass[aps,nofootinbib,twocolumn,floatfix,showpacs,preprintnumbers]{revtex4} 
\usepackage{graphicx}
\input epsf
\newcommand{\sfig}[2]{
\includegraphics[width=#2]{#1}
        }
\newcommand{\Sfig}[2]{
    \begin{figure}[thbp]
    \sfig{#1.eps}{\columnwidth}
    \caption{{\small #2}}
    \label{fig:#1}
    \end{figure}
}

\newcommand{\rf}[1]{\ref{fig:#1}}

\def\lsim{\mathrel{\raise.3ex\hbox{$<$\kern-.75em\lower1ex\hbox{$\sim$}}}}
\def\gsim{\mathrel{\raise.3ex\hbox{$>$\kern-.75em\lower1ex\hbox{$\sim$}}}}

\def\cmm2{{\,\rm cm^{-2}}}
\def\cm2{{\,{\rm cm}^2}}
\def\cmm3{{\,{\rm cm}^{-3}}}
\def\gcmm3{{\,{\rm g\,cm^{-3}}}}

\def\fun#1#2{\lower3.6pt\vbox{\baselineskip0pt\lineskip.9pt
  \ialign{$\mathsurround=0pt#1\hfil##\hfil$\crcr#2\crcr\sim\crcr}}}

\def\be{\begin{equation}}
\def\ee{\end{equation}}
\def\bea{\begin{eqnarray}}
\def\eea{\end{eqnarray}}
\newcommand{\vs}{\nonumber\\}

\newcommand{\ec}[1]{Eq.~(\ref{eq:#1})}

\newcommand{\eql}[1]{\label{eq:#1}}

\begin{document}
\preprint{FERMILAB-PUB-09-346-A-PPD, MAN/HEP/2009/31}

\title{Cosmic Neutrino Last Scattering Surface}

\author{Scott Dodelson$^{1,2,3}$,
Mika Vesterinen$^4$
}

\affiliation{$^1$Center for Particle Astrophysics, Fermi National
Accelerator Laboratory, Batavia, IL~~60510-0500, USA}
\affiliation{$^2$Department of Astronomy \& Astrophysics, The
University of Chicago, Chicago, IL~~60637-1433, USA}
\affiliation{$^3$Kavli Institute for Cosmological Physics, Chicago, IL~~60637-1433, USA}
\affiliation{$^4$The School of Physics and Astronomy, The University of
Manchester, Manchester M13 9PL, UK}
\date{\today}
\begin{abstract}
Neutrinos decoupled from the rest of the cosmic plasma when the Universe was
less than one second old, far earlier than the photons, which decoupled at
$t=380,000$ years. Surprisingly, though, the last scattering surface of massive
neutrinos is much closer to us than that of the photons. Here we calculate the
properties of the last scattering surfaces of the three species of neutrinos.
\end{abstract}
\maketitle

\section{Introduction}

The standard cosmological model predicts that neutrinos were produced in the
early universe and are present today with an abundance\footnote{The
uncertainty on this prediction is set by the sub-percent uncertainty in the temperature of the cosmic
microwave background temperature~\cite{Fixsen}, which serves to calibrate the 
thermal number density. The
predicted abundance is roughly one
percent larger than thermal due to $e^+/e^-$ heating~\cite{sdmt}.} of $112$ cm$^{-3}$ per
species~\cite{KolbTurner,Dodelson:2003ft}. Detecting this background remains a 
tantalizing experimental dream~\cite{Orpher,Stodolsky,Hagmann,Lewis,Cabibo,Smith,Langacker}, with
recent developments encouraging the
optimists~\cite{Lazauskas:2007da,Cocco,Blennow:2008fh,Volpe:2009ck,
Cocco:2009rh,McElrath:2009ig,Ringwald:2009bg}. 

For the time being, the question of where the cosmic neutrinos come from remains
an academic question. Yet it is of sufficient interest that, even if there were
no chance of detection, the origin
of the cosmic neutrino background (CNB) seems worthy of theoretical study. 

The
neutrino {\it Last Scattering Surface} (LSS) is typically thought of as being located
a given distance from us with a small but finite width, similar to the last
scattering surface of the cosmic microwave background (CMB). 
Neutrinos last scatter when the temperature of the
universe was a few MeV and the universe was less than a second old, while the
photons in the CMB last scattered much later when the temperature was $1/3$ eV
at $t=380,000$ years, so it is natural to assume that the neutrino LSS is further away than
that of the CMB. Calculating how much further away leads to a little surprise.
For, even massless particles
can travel only a very small comoving distance in the very early universe, so
the CMB comes from a comoving distance about 9540 $h^{-1}$ Mpc away from us
(where the present expansion rate is $H_0=100\, h$ km s$^{-1}$ Mpc$^{-1}$), while
{\em massless} neutrinos arrive from a comoving distance 9735 $h^{-1}$ Mpc away. That is, neutrinos
travel only about 200 Mpc (comoving) in the first 380,000 years.

We show in this paper that this slightly surprising result evolves into another
counter-intuitive result for massive neutrinos~\cite{Kogan,Kogan1}: the CNB actually reaches us from
{\it closer} than the CMB. Even for neutrino masses as small as 0.05 eV (and one of the
neutrinos must be at least this massive) the effect is dramatic. For neutrinos
with mass of 1 eV, the effect is truly striking with most of these neutrinos
arriving from only several {\it hundred} Mpc away! 

\section{Last Scattering Surface of Massive Neutrinos}

Neutrinos stopped scattering when temperatures were of order a few MeV and the 
universe was less than a second old. At that time each neutrino species had a
Fermi-Dirac distribution for a massless particle (assuming -- as we will
throughout -- zero chemical potential and no heating from electron-positron annihilation). 

Consider first the calculation of the last scattering surface of a massless
neutrino. The comoving distance travelled by a massless particle starting from
$t_i=1$ sec until today is
\be
\chi = \int_{t_i}^{t_0} \frac{dt}{a(t)}
= \int_{a_i}^1 \frac{da}{a^2 H(a)}
.\ee
Note that early on $H(a)= H_0\Omega_r^{1/2} a^{-2}$ where 
$\Omega_r=8.3\times 10^{-5}$
is the radiation density {\it today} in units of the critical density. The integrand peaks at late times,
so the 
contribution to the comoving distance from early times ($t_i\sim1$ sec)
is negligible; this explains our nonchalance in defining the initial time. It
also explains why all three species of neutrinos share the same last scattering
surface even though electron neutrinos decouple slightly later than do the other
two species. The comoving distance travelled by neutrinos
until the time of photon last scattering at $a_*^{-1}=1090.5\pm0.95$~\cite{wmap} is:
\bea
\chi_* 
&=&  \frac{1}{\Omega_m^{1/2}H_0} \int_{a_i}^{a_*} \frac{da}{\sqrt{a + a_{\rm 
eq}}}\vs
&=& \frac{2}{\Omega_m^{1/2}H_0} \left[ \sqrt{a_* + a_{\rm eq}} -\sqrt{a_{\rm eq}}
\right]
\eea
where $\Omega_m$ is the matter density today in units of the critical density and
$a_{\rm eq}$ is the value of the scale factor when the densities of matter and radiation
are equal. 
Using the standard cosmological parameters~\cite{wmap}, this equates to the 200
$h^{-1}$ Mpc difference alluded to in the introduction.

Massive neutrinos slow down once they become non-relativistic, so the integral
determining the distance to the last scattering surface generalizes to:
\be
\chi = \int_{t_i}^{t_0} \frac{dt}{a(t)} \frac{p_0/a}{\sqrt{(p_0/a)^2+m_\nu^2}}
\eql{chip}\ee
where the second term in the integrand is the redshifted velocity $p/E$, with
$p_0$ the current neutrino momentum. The neutrino temperature today is
$T_\nu=1.95\times 10^{-4}$ eV, so there will be a range of $p_0$'s drawn from
a Fermi-Dirac distribution, each of which will be associated with a different
distance to the LSS. Fig.~\rf{chim} plots this distance as a function of
neutrino mass for two different values of the present day neutrino momentum. 

\Sfig{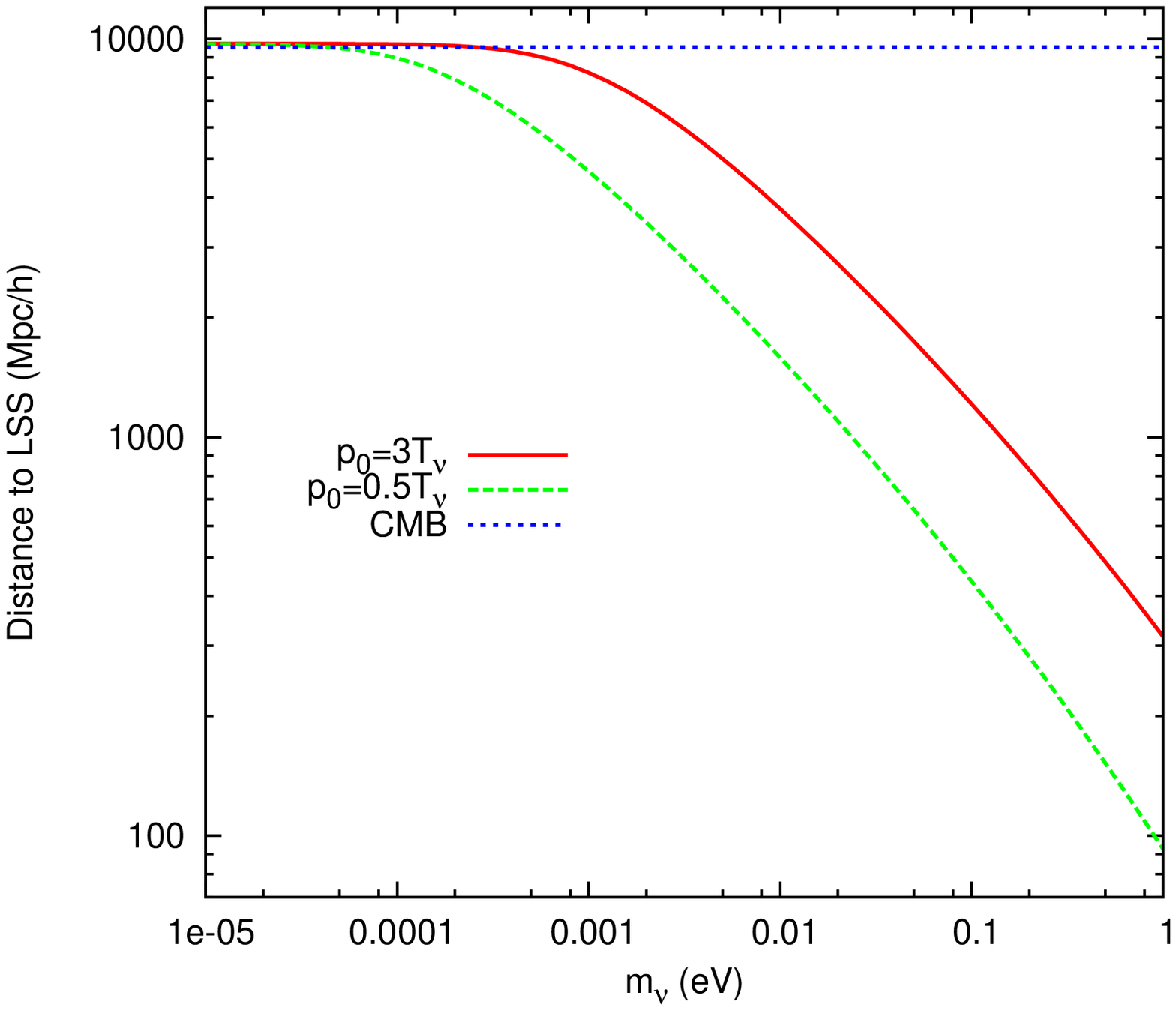}{The comoving distance travelled by a massive neutrino since
decoupling until today as a function of mass for two different values of the
neutrino momentum. Neutrinos with smaller momenta are travelling more slowly and
therefore travel a shorter distance. Note that for masses above $10^{-4}$ eV,
the neutrino LSS is much closer than that of the CMB.}

Since neutrinos with different momenta arrive from different distances\footnote{Ref.~\cite{Jia}
also mentioned this feature of neutrinos and proposed to exploit it to test the Copernican 
Principle.}, 
the last
scattering surface of the CNB is quite broad compared to that of the CMB. 
To quantify this, we can define the probability that a neutrino last scattered a
distance $\chi$ away from us, or the {\it visibility function}:
\be
\frac{dP}{d\chi} = \frac{dP}{dp_0} \left( \frac{d\chi}{dp_0} \right)^{-1}
\ee
where the equality uses the chain rule; the first differential probability
is given by the massless Fermi-Dirac distribution~\cite{Bernstein}:
\be
\frac{dP}{dp_0}= \frac{2}{3\zeta(3)T_\nu^3} \frac{p_0^2}{e^{p_0/T_\nu}  +1};
\ee
and the second term on the right is obtained by differentiating \ec{chip}.

Fig.~\rf{chinu} displays the probability that a neutrino with a given mass
arrived from a distance $\chi$. Note for all masses above $10^{-4}$ eV, the
spread in arrival distances is much larger than the spread in the CMB last
scattering surface. The width of the CNB and CMB last scattering surfaces have
different origins: the CMB does not have an infinitely thin last scattering
surface because the process of recombination, and therefore decoupling, extends
over a finite time period. The CNB last scattering surface is thick, reflecting
the different velocities of the different momenta in the Fermi-Dirac
distributions. 

\Sfig{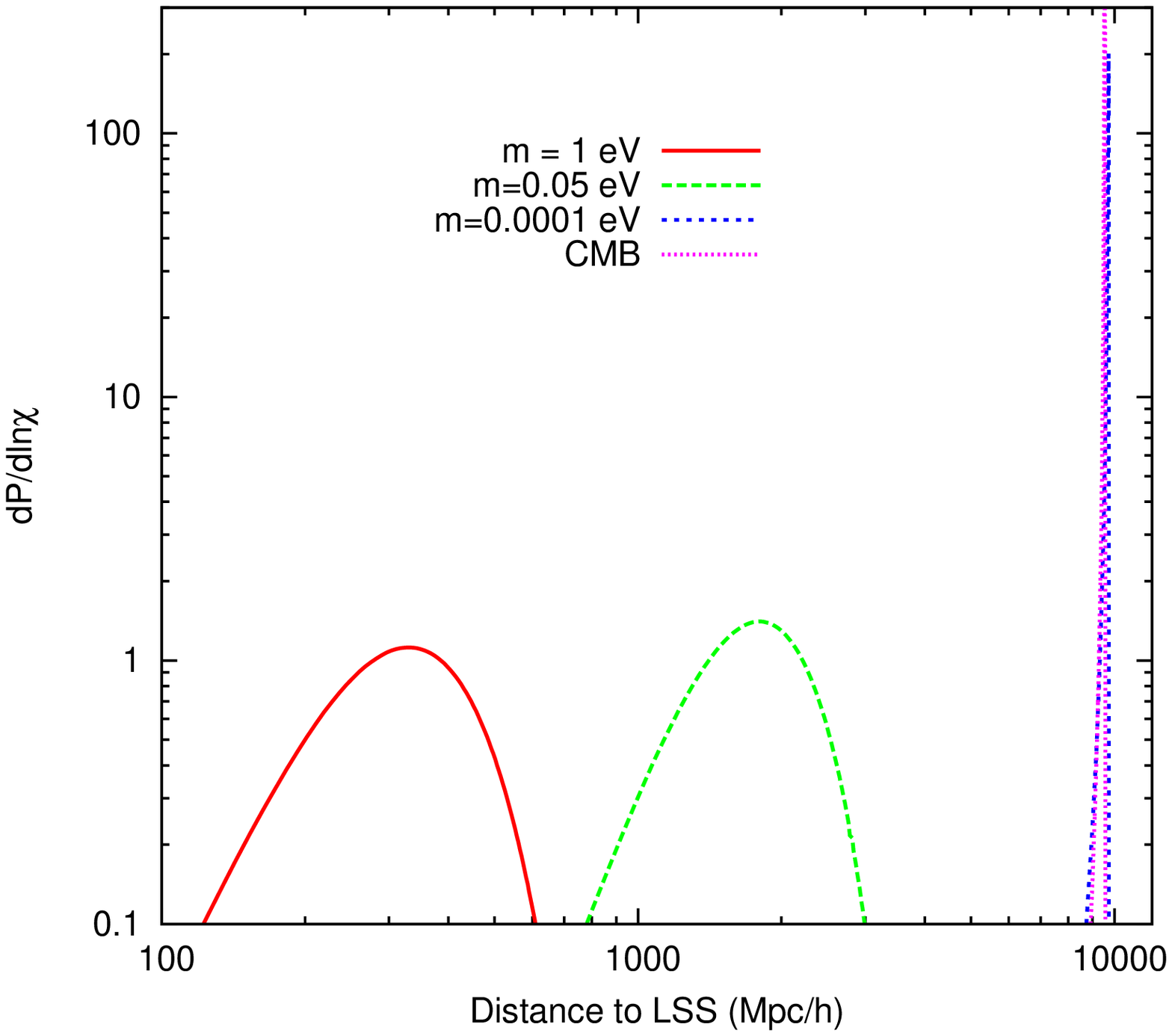}{The probability that a neutrino with mass $m$ last scatters at a
given comoving distance from us (the visibility function). 
Massive neutrinos travel more slowly than
massless neutrinos so arrive here from much closer distances. Also shown is the
last scattering surface of the cosmic microwave background, virtually
indistinguishable from that of an $m_\nu=10^{-4}$ eV neutrino.}

One might ask about the last scattering surface of heavier particles, such as sterile
neutrinos or ordinary cold dark matter particles. Those that
become non-relativistic before equality travel a distance of order the comoving horizon at
equality (136 Mpc) times the comoving velocity at equality [$(p_0/m)/a_{\rm EQ}$]. A 
keV sterile neutrino therefore would have a last scattering surface of order a Mpc away.
For particles more massive than this -- e.g., Cold Dark Matter -- their LSS is so close that
the distance traveled in a straight line (before gravity in our halo started moving
them around) was negligible. So in some sense, the question ceases to make much sense
for masses above a keV. 
Some of these considerations apply even to light neutrinos in our Galaxy: to determine
what has happened to these neutrinos recently, one must carry out simulations along  the
lines of those presented in \cite{wong}.

\section{Oscillations}

Neutrinos are produced in flavor eigenstates but then propagate as mass
eigenstates. Therefore, the simple notion of a neutrino with fixed mass having
its own last scattering surface is a bit too naive. As emphasized in
\cite{Fuller},
the number density of neutrinos in mass eigenstate $i$ is:
\be
dn_i = \sum_{\alpha=1}^3 \left\vert U_{\alpha i}\right\vert^2
dn_\alpha
\ee
where $U_{\alpha i}$ are the elements of the unitary matrix which transforms
from the mass basis to the flavor basis, and the sum is over all three flavors.

In principle the number densities of the three neutrinos flavors $dn_\alpha$
could differ if, e.g., each had a different chemical potential~\cite{Dolgov}. Here we assume
that the chemical potentials are very small so all the $dn_\alpha$ are for all practical
purposes identical. In that case, they can be removed from the sum, and then the
sum over the unitary matrix elements squared simply gives unity. So 
\be
dn_i = dn_\alpha = \frac{dp_0}{2\pi^2} \frac{p_0^2}{e^{p_0/T_\nu}  +1}
\ee
independent of mass eigenstate. So we expect the calculation of the previous
section to reflect the different last scattering surfaces of the different mass
eigenstates. 

Detection, however, will take place in flavor space, with
electron neutrinos and anti-neutrinos. To compute the last scattering surface of
these eigenstates, we must transform back to flavor space, weighting each mass
eigenstate by its own visibility function:
\be
\frac{dP}{d\chi}\big\vert_{\nu_e} = \frac{dP}{dp_0} 
\sum_i \left\vert U_{ei}\right\vert^2
\left( \frac{d\chi}{dp_0} \right)_{\nu_i}^{-1}
.\ee
Fig.~\rf{chinue} shows the resulting probability for an electron neutrino 
with two possible
mass schemes assuming the tri-bimaximal mixing matrix~\cite{Harrison1,Harrison2} ($U_{e1}=\sqrt{2/3}$,
$U_{e2}=1/\sqrt{3}$, and $U_{e3}=0$). Note the interesting double 
peaked structure in the
normal case, a signature of the quantum mechanical oscillations which dictate
that there is roughly a 2/3 probability the electron neutrino propagates with very
small mass $m_1$ and 1/3 with mass $m_2$.

\Sfig{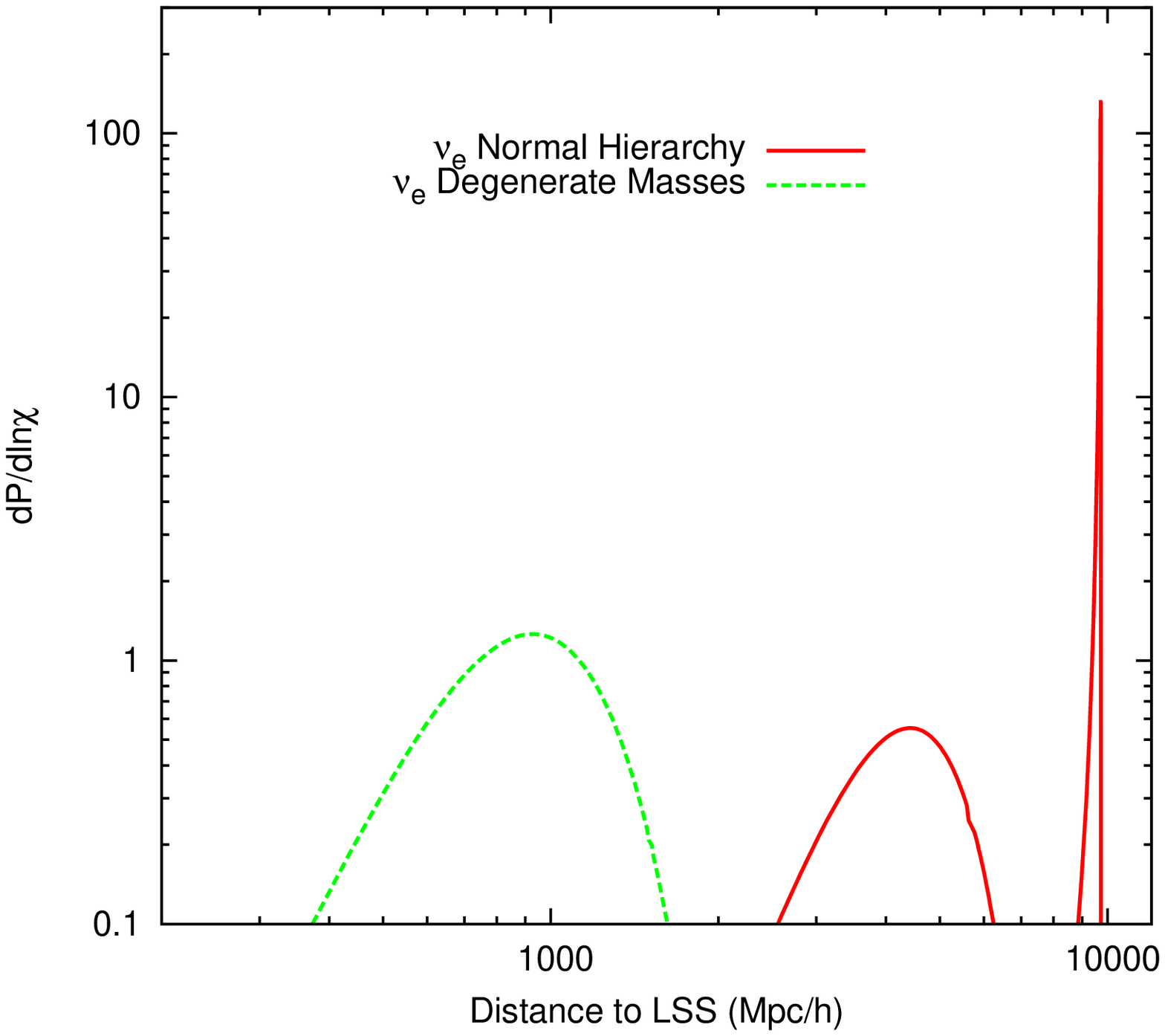}{Probability for the distance to the last scattering surface for an
electron neutrino with normal hierarchy ($m_2=0.009$ eV; $m_1=10^{-4}$ eV;
$\sin\theta_{13}=0$) and degenerate masses ($m_1=m_2=m_3=0.2$ eV).}

\section{Conclusions}

The last scattering surface of the cosmic neutrino background is much broader and much closer
than that of the cosmic microwave background. Indeed, as depicted in Fig.~\rf{chinue}, the last
scattering ``surface'' of the electron neutrinos and anti-neutrinos has a very rich structure
due to oscillations.

Are there any observable consequences of these distances to the last scattering surface? We
can think of three potentially interesting follow-ups:

\begin{itemize}

\item{\bf Neutrino Acoustic Oscillations} 

The well-known {\it Baryon Acoustic Oscillations}
(BAO) arise because the photon-baryon gas travels a distance equal to the sound horizon at
decoupling, after which the baryons stop. In real space, this leads to an overdensity of
baryons in a spherical shell surrounding an initial overdensity, and ultimately to a bump
in the correlation function. Massive neutrinos would seem to share many of the same
features: adiabatic perturbations lead to neutrinos initially being overdense at the 
same places as the baryons; neutrinos travel a finite distance since last scattering; and
this distance might show up as a feature in the power spectrum. Unfortunately, we see no
evidence for this feature when running the Boltzmann code~\cite{cmbfast} which solves for the linear
evolution of perturbations (and by tracking different neutrino momenta implicitly 
encodes all of the relevant physics). The difference between neutrinos and the baryons are that the
neutrino last scattering surface is very broad due to the Fermi-Dirac distribution; this
breadth smooths out any (very small) feature rendering it undetectable. Searching for a feature
might still be useful as a way of constraining the distribution function of neutrinos in the
CNB.

\item{\bf Galactic Distribution function} 

For direct detection, it is important to know the
distribution function of neutrinos in a Milky-Way sized galaxy. Are neutrinos overdense? Has
their momentum distribution changed due to virialization? If capture has occurred, the distance
to the LSS would likewise be modified. Indeed, it is possible that there would be a directional
effect: neutrinos from the center of the Galaxy might be captured and hence have relatively
small distances to the LSS, while those from the poles maintaining their primordial distribution
functions and distances to LSS.

\item{\bf Anisotropies} 

If experimentalists do succeed in detecting the
CNB, in the far future one might imagine maps of the anisotropy of this background~\cite{caldwell}.
If the neutrinos have been sloshing in our Galaxy for a number of orbits, this angular information
might be lost. But some of the neutrinos -- the lightest ones or those arriving away from
the Galactic Center -- will arrive undeflected (see, e.g., the right panel in Fig. 2 
in Ref.~\cite{Singh} which suggests
very little trapping of neutrinos with masses less than 1 eV in a Galaxy of our size). These would 
provide information about overdensities at a very early time
from locations about which we will already have much information from galaxy surveys. The galaxy surveys
probe the same locations but at much later times. This would be an almost unique opportunity to probe
the evolution of structure in given regions.

\end{itemize}

This work has been supported by the US Department of Energy, including grant DE-FG02-95ER40896.
We thank Alex Kusenko for very fruitful discussions, Wayne Hu for insight into
the neutrino effects on the power spectrum, and Nicole Bell for her neutrino expertise. We are grateful to the CTEQ Summer School, where a question at a late night
recitation led to the idea for this work, and to the Aspen Center for Physics where the work
was carried out.

\bibliographystyle{h-physrev4}
\bibliography{neutrino}
\end{document}